\documentclass[aip,jcp,reprint,floatfix]{revtex4-2}
\usepackage[T1]{fontenc}
\usepackage[utf8]{inputenc}
\usepackage{calc}
\usepackage{textcomp}
\usepackage{amsmath}
\usepackage{amssymb}
\usepackage[mathlines]{lineno}
\usepackage{graphicx}
\usepackage{hyperref}



\usepackage{babel}
\begin{document}
\title{Defect Organization in Coexisting Hexagonal and Square Lattices on Ellipsoids}

\author{Wenyu Liu}
\affiliation{School of Physics and Key Laboratory of Functional Polymer Materials of Ministry of Education, Nankai University, and Collaborative Innovation Center of Chemical Science and Engineering, Tianjin 300071, China}
\author{Han Xie}
\affiliation{Department of Physics and Astronomy, University of Waterloo, Waterloo, Ontario N2L 3G1, Canada}
\affiliation{Department of Physics, The Hong Kong University of Science and Technology, Clear Water Bay, Kowloon, Hong Kong, China}
\author{Baohui Li}
\email{baohui@nankai.edu.cn}
\affiliation{School of Physics and Key Laboratory of Functional Polymer Materials of Ministry of Education, Nankai University, and Collaborative Innovation Center of Chemical Science and Engineering, Tianjin 300071, China}
\author{Jeff Z.Y. Chen}
\email{jeffchen@uwaterloo.ca}
\affiliation{Department of Physics and Astronomy, University of Waterloo, Waterloo, Ontario N2L 3G1, Canada}
\author{Yao Li}
\email{liyao@nankai.edu.cn}
\affiliation{School of Physics and Key Laboratory of Functional Polymer Materials of Ministry of Education, Nankai University, and Collaborative Innovation Center of Chemical Science and Engineering, Tianjin 300071, China}

\date{\today}

\begin{abstract}
Curvature and topology jointly organize defects in two-dimensional crystals, but their combined role remains unresolved when competing lattice symmetries coexist with spatially varying curvature. We use simulated-annealing Langevin dynamics to study Hertzian particles forming coexisting hexagonal (Hex) and square (Sq) lattices on prolate and oblate ellipsoids. Mapping reduced density and aspect ratio reveals a broad sequence of scar and domain-based morphologies in both Hex-dominant and Sq-dominant backgrounds. Latitude-resolved comparisons show that Gaussian curvature biases defects toward its maxima under weak deformation. Strong prolateness, however, confines high curvature to small polar caps that cannot independently accommodate all defect motifs. Defects then spread toward lower-curvature latitudes to relieve defect crowding and elastic repulsion. In the Hex-dominant regime, this competition drives vertex-contacted domains with neutralized corner contacts, and compensating positive defects locate away from the poles. The Sq-dominant regime features Hex-rich triangular domains, bridged states, and linear or open scars similarly reorganized by curvature anisotropy. On oblate ellipsoids, the extended equatorial high-curvature belt allows defects to separate azimuthally while remaining curvature-localized. This work elucidates that nonuniform curvature can engineer rich defect patterns by selecting the spatial distribution of topological charge and the connectivity of finite defect motifs.

\end{abstract}

\maketitle

\section{\label{sec:level1}Introduction}

The organization of densely packed particles on curved surfaces stems
from the interplay between crystallographic order and geometric
frustration, a recurring theme in soft condensed matter physics,
materials science, and biological physics~\cite{Bowick2009,Grason2016,
Ardavseva2022}. Unlike flat two-dimensional substrates, curved manifolds
cannot sustain defect-free periodic order across the entire surface~\cite{Bowick2000interacting,Vitelli2006,Turner2010,araki2011memory,araki2013defect}.
The resulting incompatibility between local packing preferences and
global geometry yields a wide range of defect motifs, including isolated
disclinations and grain-boundary scars~\cite{Bowick2002,Bausch2003,
Irvine2010,Irvine2012}.

Spherical assemblies provide the most familiar example: Euler's theorem fixes the net topological charge, classically illustrated by the
twelve positive disclinations required for hexagonal order on a closed
surface~\cite{Bowick2002,Bausch2003,Bowick2009}. While early studies focused on
systems with a single dominant lattice symmetry, increasing attention
has turned to curved systems with nonhexagonal or competing local
order~\cite{Li2013,Manyuhina2015,Jones2026,xie2025}. In such systems, the coupling
of lattice symmetry, curvature, and topology produces defect
morphologies richer than those found in single-phase crystals
~\cite{Grason2016,Ardavseva2022,xie2025}.

Our previous simulations showed that particles interacting through the Hertzian potential provide a
convenient minimal model for this problem because both hexagonal (Hex)
and square (Sq) lattices occur over adjacent density ranges~\cite{Terao2013,Fomin2018,xie2025}. The Hertzian potential is a finite-range, purely repulsive soft-contact interaction whose energy increases continuously with particle overlap and vanishes when the particles no longer overlap. Near the solid-solid transition between
these lattice types, spherical confinement produces a variety of
coexistence patterns, most notably domain-like structures
~\cite{xie2025,xie2025filling,liu2025transitional}. These results indicate that topology enforces a nonzero net defect charge, whereas curvature reshapes the morphology and spatial distribution of coexisting lattice order
~\cite{xie2025,xie2025filling,liu2025transitional}.

Ellipsoidal confinement introduces an additional level of geometric control because the Gaussian curvature varies across the surface~\cite{Bowick2000interacting,giomi2007crystalline,Irvine2010,Vitelli2006,LopezJimenez2016}. Prolate and oblate ellipsoids exhibit distinct curvature extrema and anisotropic curvature gradients along the symmetry axis, which can influence lattice ordering~\cite{Burke2015,LopezJimenez2016}. Related studies have shown that surface geometry and interaction design can affect lattice organization and defect structures on curved interfaces and shells~\cite{Ershov2013,Agarwal2020,zhu2024programmable}. However, how curvature heterogeneity reorganizes coexisting lattice symmetries and their associated defect motifs on ellipsoids remains unclear.

Here we investigate how density and ellipsoidal anisotropy reorganize
the defect motifs associated with competing Hex and Sq order. We first
map the state diagram and then examine scar- and domain-based structures
in the Hex- and Sq-dominant regimes. Latitude-resolved comparisons with
Gaussian curvature are used to identify the geometric and elastic
mechanisms that control defect placement and domain contact.
This comparison separates two geometric responses that are absent on a uniformly curved sphere: a location response, in which intact motifs move relative to the curvature extrema, and a connectivity response, in which domains contact or bridge and thereby reorganize their local topological charges.

\section{Model and Methods}

\subsection{Model and annealing protocol}

We used simulated-annealing Langevin dynamics~\cite{Kirkpatrick1983}, implemented in LAMMPS~\cite{LAMMPS}, to search for low-energy configurations of $N=1000$ particles constrained to an ellipsoidal surface. The particles interacted through the Hertzian potential
\begin{equation}
U(r)=
\begin{cases}
\epsilon\left(1-r/\sigma\right)^{5/2}, & r<\sigma,\\
0, & r\geq\sigma,
\end{cases}
\label{eq1}
\end{equation}
where $r$ is the center-to-center distance, $\sigma$ is the particle diameter, and $\epsilon$ sets the energy scale~\cite{Terao2013,Fomin2018}. Reduced units were defined by $\sigma$, $\epsilon$, and the particle mass $m$, giving the time unit
$\tau=\sqrt{m\sigma^2/\epsilon}$.
The integration time step was $\Delta t=0.5\tau$.

Independent initial configurations were generated on the prescribed ellipsoidal surface at
$k_{\rm B}T_0/\epsilon=10^{-2}$.
The surface constraint was enforced using the RATTLE algorithm~\cite{Andersen1983,Paquay2016}, with a convergence tolerance of $10^{-4}$ and a maximum of 20 iterations per time step. Each initial configuration was equilibrated for $10^6$ steps at $T_0$. The temperature was subsequently reduced according to
\begin{equation}
T_n=T_0\exp(-0.002n),
\qquad n=0,\ldots,2499,
\label{eq2}
\end{equation}
with $10^4$ integration steps performed at each temperature. The Langevin damping time was $10\Delta t$. Each annealing trajectory was followed by a zero-temperature energy minimization.

For each state point, we performed 20 independent annealing trajectories and ranked the resulting configurations by their final potential energies. The lowest-energy configuration was used in the morphology diagram, whereas the ten lowest-energy configurations from the same set of 20 trajectories were retained for the latitude-resolved statistical analysis.

These selected configurations arise from independent annealing trajectories rather than from consecutive frames of a single trajectory.

\subsection{Ellipsoidal geometry}

The ellipsoidal surface was parameterized as
\begin{equation}
\mathbf{r}(\theta,\varphi)=
\left(
B\sin\theta\cos\varphi,\,
B\sin\theta\sin\varphi,\,
A\cos\theta
\right),
\label{eq3}
\end{equation}
where $A$ and $B$ are the polar and equatorial semiaxes, respectively, $0\leq\theta\leq\pi$ is the parametric colatitude, and $0\leq\varphi<2\pi$. Thus,
\begin{equation}
\frac{x^2+y^2}{B^2}+\frac{z^2}{A^2}=1.
\label{eq4}
\end{equation}
The cases $A/B>1$, $A/B=1$, and $A/B<1$ correspond to prolate ellipsoids, a sphere, and oblate ellipsoids, respectively.

The dimensionless surface density was
\begin{equation}
\rho^*=\frac{N\sigma^2}{S},
\label{eq5}
\end{equation}
where $S$ is the total ellipsoidal area. For each prescribed pair of $\rho^*$ and $A/B$, the semiaxes were rescaled such that $S=N\sigma^2/\rho^*$. The spheroidal area was evaluated analytically as
\begin{equation}
S=
\begin{cases}
2\pi B^2\left(1+\dfrac{A}{Be}\sin^{-1}e\right),
& A>B,\quad e=\sqrt{1-\dfrac{B^2}{A^2}},\\[8pt]
4\pi A^2,
& A=B,\\[6pt]
2\pi B^2\left(1+\dfrac{A^2}{B^2e}\operatorname{artanh}e\right),
& A<B,\quad e=\sqrt{1-\dfrac{A^2}{B^2}}.
\end{cases}
\label{eq6}
\end{equation}

For the above parameterization, the Gaussian curvature is
\begin{equation}
K(\theta)=
\frac{A^2}
{\left(A^2\sin^2\theta+B^2\cos^2\theta\right)^2},
\label{eq7}
\end{equation}
and the surface-area element integrated over the azimuthal angle is
\begin{equation}
{\rm d}A=
2\pi B\sin\theta
\sqrt{A^2\sin^2\theta+B^2\cos^2\theta}\,
{\rm d}\theta.
\label{eq8}
\end{equation}
These expressions were used both to calculate the curvature profile and to determine the exact area of each latitude bin~\cite{DoCarmo1976}.

\subsection{Local structure and defect identification}

Local structures were identified from the bond network between
neighboring particles. Two particles were defined as bonded when their center-to-center separation $r_{ij}$ satisfied $r_{ij}<\sigma$. Closed loops in the resulting bond network were classified by their number of boundary bonds: a loop bounded by three bonds was identified as a triangular cell, whereas a loop bounded by four bonds was identified as a quadrilateral cell. These cells were then used to distinguish the local Hex and Sq environments.

Particles surrounded by both types of cells were assigned to the Hex-Sq interface.

For particle $j$, let $n_j^{(3)}$ and $n_j^{(4)}$ denote the numbers of triangular and quadrilateral cells incident on it, respectively. Its winding number was calculated as
\begin{equation}
w_j=
1-\frac{n_j^{(3)}}{6}
-\frac{n_j^{(4)}}{4}.
\label{eq9}
\end{equation}
In a purely Hex environment, this expression reduces to
$w_j=1-Z_j/6$, where $Z_j$ is the number of bonded neighbors. In a purely Sq environment, it reduces to $w_j=1-Z_j/4$. As a consistency check, the identified defects satisfy the Euler constraint
\begin{equation}
\sum_j w_j=2
\label{eq10}
\end{equation}
on the closed ellipsoidal surface.

Here the winding number is the dimensionless local topological charge. The corresponding angular deficit is

\begin{equation*}
q_j=2\pi w_j
=2\pi-n_j^{(3)}\frac{\pi}{3}-n_j^{(4)}\frac{\pi}{2}.
\end{equation*}
For example, fivefold and sevenfold sites in a purely Hex environment carry winding numbers $+1/6$ and $-1/6$, respectively.

\subsection{Latitude-resolved defect-curvature comparison}

The surface was divided into nonoverlapping latitude bins of width
$\Delta\theta=5^\circ$, spanning $\theta=0$ to $\theta=\pi$, with each bin labeled by its central latitude $\theta$. The area associated with each latitude bin was
\begin{equation}
A(\theta)=
\int_{\theta-\Delta\theta/2}^{\theta+\Delta\theta/2}{\rm d}A,
\label{eq11}
\end{equation}
and the corresponding area-averaged Gaussian curvature was
\begin{equation}
K(\theta)=
\frac{1}{A(\theta)}
\int_{\theta-\Delta\theta/2}^{\theta+\Delta\theta/2}
K(\theta')\,{\rm d}A.
\label{eq12}
\end{equation}
The signed topological-charge surface density was defined as
\begin{equation}
\rho(\theta)=
\frac{1}{A(\theta)}
\sum_{j\in\theta} w_j,
\label{eq13}
\end{equation}
where the sum runs over particles within the corresponding latitude bin. For each latitude bin, positive and negative winding numbers were summed algebraically, so negative winding numbers reduce $\rho(\theta)$. Consequently, $\rho(\theta)$ is a signed topological-charge density rather than a positive-defect number density. The normalized signed topological-charge density and curvature profiles were then defined as
\begin{equation}
\widetilde{\rho}(\theta)=
\frac{\rho(\theta)}{\langle\rho\rangle},
\qquad
\widetilde{K}(\theta)=
\frac{K(\theta)}{\langle K\rangle},
\label{eq14}
\end{equation}
where
\begin{equation}
\langle\rho\rangle=
\frac{W}{S},
\qquad
\langle K\rangle=
\frac{1}{S}\int K\,{\rm d}A
=\frac{4\pi}{S},
\label{eq15}
\end{equation}
and positive and negative winding numbers enter the normalization algebraically. Because the latitude bins cover the complete closed surface, $W=\sum_j w_j=2$, consistent with the Euler constraint in Eq.~\ref{eq10}.

The mismatch between the two profiles was measured using
\begin{equation}
D_{\rho K}
=
\frac{(1-r_{\rho K})\sigma_K}
{2\langle K\rangle},
\label{eq16}
\end{equation}
where $r_{\rho K}$ is the Pearson correlation coefficient between the 36 latitude-bin values of $\rho(\theta)$ and $K(\theta)$, and $\sigma_K$ is the unweighted standard deviation of the corresponding 36 values of $K(\theta)$. The correlation coefficient and $D_{\rho K}$ were calculated separately for each realization and then averaged over the ten lowest-energy configurations selected from the same 20 independent annealing trajectories. Error bars and shaded regions represent the standard error of the mean.

Particle configurations were visualized using OVITO~\cite{OVITO}.

\section{\label{sec:level1-2}Results and Discussion}

\begin{figure}[htbp]
\begin{centering}
\includegraphics[width=1\columnwidth]{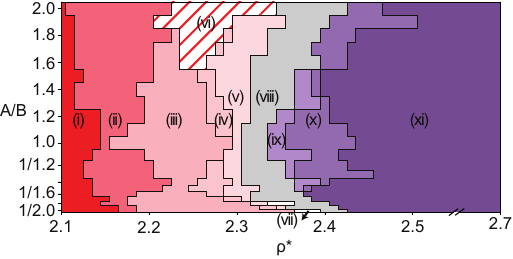}
\par\end{centering}
\centering{}\caption{State diagram of mixed Hex-Sq lattices on ellipsoidal surfaces as
functions of reduced density $\rho^*$ and aspect ratio $A/B$. Red,
purple, and gray backgrounds denote the Hex-dominant, Sq-dominant, and
intermediate regimes, respectively. Roman numerals identify
(i) Hex scars (see Fig.~\ref{Fig2}), (ii) coexistence of Hex scars and Sq-rich rectangular domains (see Fig.~\ref{Fig7}),
(iii) Sq-rich rectangular domains (see Fig.~\ref{Fig4}),
(iv) coexistence of Sq-rich rectangular domains and counter domains (see Fig.~\ref{Fig9}),
(v) counter domains (see Fig.~\ref{Fig8}), (vi) vertex-contacted domains (see Fig.~\ref{Fig6}),
(vii) biangular domains (see Fig.~\ref{Fig10}), (viii) asymmetric mixed state,
(ix) Hex-rich triangular domains (see Fig.~\ref{Fig11}), (x) bridged states (see Fig.~\ref{Fig13}),
and (xi) linear and open scars (see Fig.~\ref{Fig14}).
Each state point represents the lowest-energy configuration found
among 20 independent annealing runs. The red hatching is used solely to draw attention to the novel VCD region (vi) and does not denote an additional phase, coexistence, bistability, insufficient sampling, or uncertainty in a phase boundary.}
\label{Fig1}
\end{figure}

\subsection{State diagram}

Figure~\ref{Fig1} summarizes the lowest-energy configurations found over the $(\rho^*,A/B)$ parameter space. At each state point, we performed 20 independent annealing runs and selected the configuration with the lowest final energy. We used $\Delta \rho^*=0.01$, aspect ratios $A/B=1.1,\,1.2,\ldots,2.0$ for prolate ellipsoids, and the reciprocal series $A/B=1/2.0,\,1/1.9,\ldots,1/1.1$ for oblate ellipsoids.

The diagram contains three broad regimes. Lower densities generally favor Hex-dominant configurations, higher densities favor Sq-dominant configurations, and intermediate morphologies occur near the crossover between them. The precise boundaries depend on the aspect ratio $A/B$.
The mixed states have lower global symmetry and stronger phase separation than the states on either side.

\begin{figure}[htbp]
\begin{centering}
\includegraphics[width=1\columnwidth]{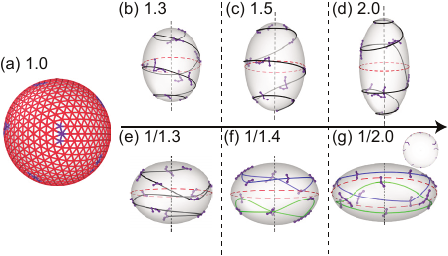}
\par\end{centering}
\centering{}\caption{\label{Fig2}Representative Hex-scar configurations corresponding to state (i) in Fig.~\ref{Fig1} at $\rho^*=2.10$.
(a) Spherical reference. (b)--(d) Prolate ellipsoids with
$A/B=1.3$, $1.5$, and $2.0$. (e)--(g) Oblate ellipsoids with
$A/B=1/1.3$, $1/1.4$, and $1/2.0$. Purple particles mark non-sixfold-coordinated particles. Black, blue, and green curves are visual guides connecting scar centers. The black dashed line connects the two poles, while the red dashed line marks the equatorial plane. The arrow indicates increasing deviation from sphericity, corresponding to increasing prolateness in the upper row and increasing oblateness in the lower row.}
\end{figure}

Ellipsoidal confinement preserves the main spherical motifs but changes their placement and connectivity. The Hex-dominant states include (i) Hex scars [Fig.~\ref{Fig2}], (ii) scar-domain coexistence [Fig.~\ref{Fig7}], (iii) Sq-rich rectangular domains [Fig.~\ref{Fig4}], (iv) domain--counter-domain coexistence [Fig.~\ref{Fig9}], (v) counter domains [Fig.~\ref{Fig8}], (vi) vertex-contacted domains [Fig.~\ref{Fig6}], and (vii) biangular domains [Fig.~\ref{Fig10}]. The Sq-dominant states include (ix) Hex-rich triangular domains [Fig.~\ref{Fig11}], (x) bridged states [Fig.~\ref{Fig13}], and (xi) linear and open scars [Fig.~\ref{Fig14}]. The following sections focus on the geometric rules that organize these motifs. We omit the disordered (viii) asymmetric mixed state, characterized by maze-like intertwining without global symmetry\cite{xie2025}, because it lacks a well-defined topological motif for systematic classification.

\subsection{Defect patterns in a Hex-dominant background}

\textit{Hex scars.}---Figure~\ref{Fig2} shows the Hex-scar states at $\rho^{*}=2.10$. In a Hex-dominant background, we use ``Hex scar'' to denote a grain-boundary chain composed mainly of alternating five- and sevenfold coordinated sites and carrying a net positive winding number. On a sphere, the twelve net-positive scar complexes adopt an approximately icosahedral arrangement [Fig.~\ref{Fig2}(a)]~\cite{Bausch2003}.

Deforming the sphere breaks this global symmetry. On prolate ellipsoids, the scars form a helical sequence around the long axis. Increasing $A/B$ stretches this sequence but does not change its basic organization. On weakly oblate ellipsoids, the helical pattern remains. For stronger oblateness, $A/B\leq1/1.4$, the scars form two nearly symmetric rings on opposite sides of the equator. Their projection along the polar axis resembles the twelve marks of a clock [upper right inset of Fig.~\ref{Fig2}(g)].

\begin{figure}[htbp]
\begin{centering}
\includegraphics[width=1\linewidth]{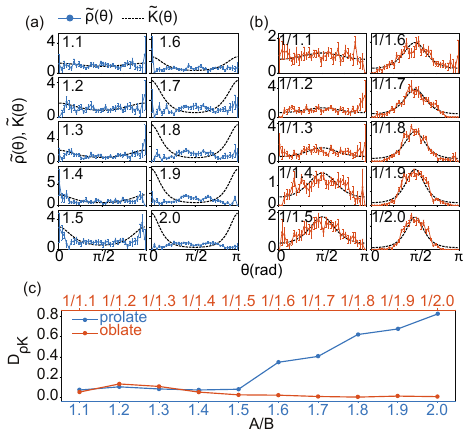}
\par\end{centering}
\centering{}\caption{\label{Fig3}Latitude-resolved curvature comparison for scars in the Hex-dominant
regime at $\rho^*=2.10$. Panels (a) and (b) show the normalized
signed topological-charge density $\widetilde{\rho}(\theta)$ for prolate and oblate
ellipsoids, respectively; dashed curves show the normalized Gaussian
curvature $\widetilde{K}(\theta)$. Panel (c) shows the defect-curvature
deviation index $D_{\rho K}$. Symbols and error bars denote the mean and
standard error over the ten lowest-energy configurations selected from the 20 independent annealing trajectories at each state point.}
\end{figure}

The helical and ring-like patterns have a common origin. Rotational symmetry makes all longitudes equivalent, so defect repulsion favors azimuthal separation. Gaussian curvature varies with latitude and biases the positive charge toward the curvature maxima. Combining azimuthal separation with curvature-selected latitudes produces a helix on a prolate ellipsoid and rings near the equator of an oblate ellipsoid.

Figure~\ref{Fig3} quantifies this relation using the latitude-resolved profiles defined in Eq.~\ref{eq14}. We compare the normalized signed topological-charge density,
$\widetilde{\rho}(\theta)=\rho(\theta)/
\langle\rho\rangle$, with the normalized Gaussian curvature,
$\widetilde{K}(\theta)=K(\theta)/\langle K\rangle$, in latitude-resolved surface bins. Normalization by the respective surface averages allows the comparison to probe spatial co-localization rather than absolute magnitude.

For weakly deformed prolate ellipsoids, the high-curvature polar regions remain sufficiently broad to accommodate both curvature-driven scar localization and scar-scar separation. The two profiles are therefore similar. As $A/B$ increases, the Gaussian curvature becomes concentrated within progressively smaller polar caps. Although these caps attract the net-positive scar complexes, confining all scars within them would increase crowding and mutual elastic repulsion. Therefore, some scars are found near the relatively low-curvature equator, causing the two profiles to separate rapidly for $A/B\gtrsim1.6$. We quantify this mismatch using
$D_{\rho K}=(1-r_{\rho K})\sigma_K/[2\langle K\rangle]$, where
$r_{\rho K}$ is the Pearson correlation coefficient between the two profiles and $\sigma_K$ is the standard deviation of $K(\theta)$.

On oblate ellipsoids, the high-curvature region forms an extended equatorial belt rather than two compact caps. This belt provides sufficient azimuthal space for scars to separate while remaining close to the curvature maximum. Consequently, $D_{\rho K}$ remains small throughout the oblate series. Thus, topology fixes the net charge but not its local distribution; the observed scar profile reflects competition among curvature-driven localization, defect repulsion, and the available area of the high-curvature region.

\begin{figure}[htbp]
\begin{centering}
\includegraphics[width=1\columnwidth]{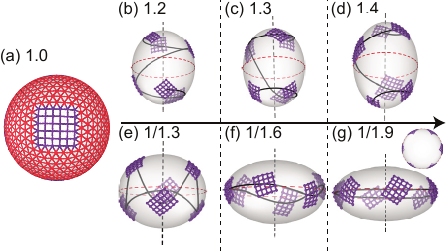}
\par\end{centering}
\centering{}\caption{\label{Fig4}Representative Sq-rich rectangular domains in a Hex-dominant background corresponding to state (iii) in Fig.~\ref{Fig1} at $\rho^*=2.24$ for several aspect ratios $A/B$.
(a) Approximately octahedral arrangement on the sphere.
(b)--(d) Prolate configurations with $A/B$ = 1.2, 1.3, and 1.4, respectively, in which the six domains form an
increasingly extended polar helix. (e)--(g) Oblate configurations with $A/B$ = 1/1.3, 1/1.6, and 1/1.9, respectively, in
which the domains approach and eventually occupy the equatorial belt.
Black curves are visual guides connecting the domain centers.
Aspect ratios are indicated in the individual panels.}
\end{figure}

\textit{Sq-rich rectangular domains.}---Near the Hex-Sq transition, Sq-rich rectangular domains form within the Hex-dominant background~\cite{xie2025}. Each of the four domain vertices carries charge $+1/12$, giving a total charge of $+1/3$ per domain. On the sphere, six such domains form an approximately octahedral arrangement [Fig.~\ref{Fig4}(a)].

On prolate ellipsoids, the six domains form a helix around the long axis [Figs.~\ref{Fig4}(b)--(d)]. Curvature draws the domains toward the poles, while their finite size and mutual repulsion maintain azimuthal separation. Increasing $A/B$ stretches the helix and intensifies crowding within the polar caps.

Oblate ellipsoids instead support a single domain ring [Figs.~\ref{Fig4}(e)--(g)]. As the ellipsoid flattens, the domain ring progressively approaches the equator and, for $A/B\lesssim1/1.9$, becomes fully confined to the equatorial region [Fig.~\ref{Fig4}(g)]. The extended high-curvature belt provides enough azimuthal space for all six domains to remain separate.

\begin{figure}[htbp]
\begin{centering}
\includegraphics[width=1\linewidth]{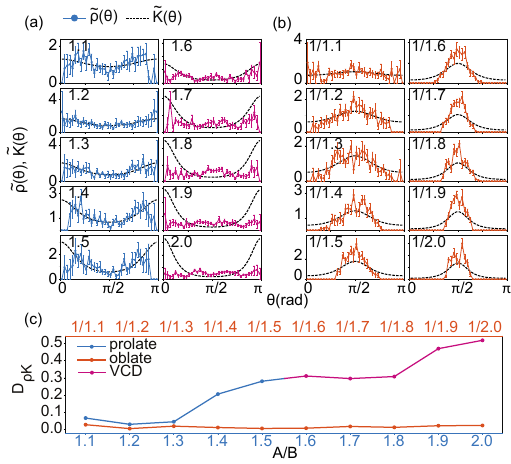}
\par\end{centering}
\centering{}\caption{\label{Fig5}
Latitude-resolved comparison of the normalized defect-charge density
$\widetilde{\rho}(\theta)$ and normalized Gaussian curvature
$\widetilde{K}(\theta)$ for Sq-rich rectangular domains in the
Hex-dominant regime at $\rho^*=2.24$.
(a) Latitude-dependent profiles for prolate ellipsoids. Blue symbols denote
the prolate series with $A/B\leq1.5$, while magenta symbols denote the VCD
states with $A/B\geq1.6$.
(b) Corresponding profiles for oblate ellipsoids, shown by orange symbols.
(c) Defect-curvature deviation index $D_{\rho K}$ as a function of $A/B$.
}
\end{figure}

The latitude-resolved profiles in Fig.~\ref{Fig5} show that
$\widetilde{\rho}(\theta)$ increasingly departs from
$\widetilde{K}(\theta)$ once $A/B$ exceeds approximately 1.3. As the prolate ellipsoid becomes more elongated, Gaussian curvature becomes increasingly localized within the polar caps, while the finite-sized domains cannot remain confined to these narrow high-curvature regions without substantial crowding. The positive defects therefore spread progressively toward lower-curvature latitudes. With further increasing $A/B$, the polar caps eventually become too confined to accommodate the domains while maintaining their separation. At $A/B\geq1.6$, neighboring domains near the poles begin to contact at their vertices, marking a change in domain connectivity. We refer to these structures as vertex-contacted-domain (VCD) states.

\begin{figure}[htbp]
\begin{centering}
\includegraphics[width=1\columnwidth]{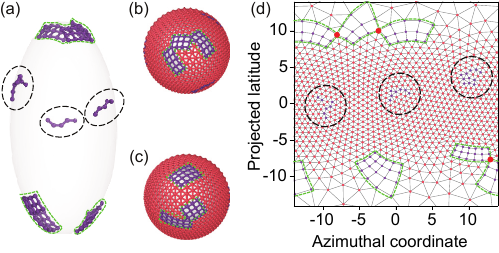}
\par\end{centering}
\caption{Vertex-contacted domains (VCD) on prolate ellipsoids corresponding to state (vi) in Fig.~\ref{Fig1}. (a) Representative
side views and an enlarged polar contact region. (b) and (c) Polar
views of the contacted domains. (d) Planar projection of the ellipsoidal lattice. Red dots mark locally neutral contacted vertices, and
black dashed outlines identify compensating positive defects at
lower-curvature latitudes. In the large side views, the defect-free Hex background is omitted and only the defect structures are shown; the accompanying top- and bottom-view insets, where present, retain the complete surface lattice. The same visualization convention is used for all subsequent snapshots.}
\label{Fig6}
\end{figure}

\textit{Vertex-contacted domains.}---Figure~\ref{Fig6} shows representative VCD configurations at $\rho^*=2.24$, including the local structures formed at the contacted vertices.

An isolated domain vertex is surrounded by one quadrilateral cell and four triangular cells and carries $w=+1/12$. As highlighted by the red dots in Fig.~\ref{Fig6}, contact reconstructs the local environment into two quadrilaterals and three triangles, for which $w=0$. Each contact therefore locally neutralizes two $+1/12$ corner charges from the polar region. Because topology fixes the net charge on the closed surface, additional positive defects appear at lower-curvature latitudes, as marked by the black dashed lines in Fig.~\ref{Fig6}.

Consistent with the VCD reconstruction, the deviation between the signed defect-charge and curvature profiles evolves nonmonotonically with increasing prolateness. For $1.6\leq A/B\leq1.8$, the occurrence of vertex contacts and compensating defects coincides with a short plateau in the deviation index. This behavior indicates that vertex sharing partially relaxes the curvature-driven localization of the domains, allowing the defect structure to reorganize without placing all positive vertices independently at the polar curvature maxima. When $A/B$ exceeds approximately $1.9$, however, the deviation increases again. Thus, vertex contact delays the separation between the defect and curvature distributions but cannot prevent it once the curvature inhomogeneity becomes sufficiently strong. By contrast, on oblate ellipsoids, the domains remain separated along the high-curvature belt, and the signed defect-charge profile continues to follow the curvature profile more closely.

\begin{figure}[htbp]
\begin{centering}
\includegraphics[width=1\columnwidth]{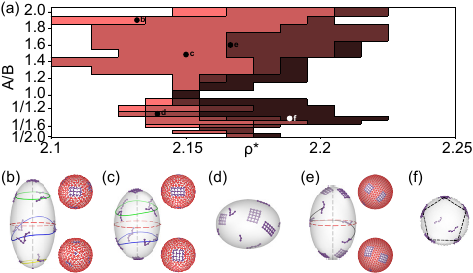}
\par\end{centering}
\centering{}\caption{\label{Fig7}Coexistence of Hex scars and Sq-rich rectangular domains corresponding to state (ii) in Fig.~\ref{Fig1}. (a) Detailed view of the state-(ii) region in Fig.~\ref{Fig1}, where Hex scars and Sq-rich rectangular domains coexist. (b)--(f) Representative
configurations containing, respectively, 10 scars and 1 domain,
8 scars and 2 domains, 6 scars and 3 domains, 4 scars and 4 domains,
and 2 scars and 5 domains.}
\end{figure}

\textit{Coexistence of Hex scars and Sq-rich rectangular domains.}---Between the scar-only and domain-only limits, Fig.~\ref{Fig7} reveals a coexistence regime in which scars and domains share the required topological charge, with Fig.~\ref{Fig7}(a) providing a detailed view of the corresponding state-(ii) region in Fig.~\ref{Fig1}. With increasing $\rho^*$, the number of scars decreases while the number of domains increases. Density therefore controls the scar-to-domain conversion, whereas ellipsoidal anisotropy determines the observed placement of the two defect motifs.

This spatial selection is most apparent on prolate ellipsoids. Domains are
preferentially trapped by the high-curvature polar regions, while scars are
pushed toward lower-curvature latitudes. In the ``10 scars + 1 domain'' state,
which occurs mainly at low $\rho^*$ and large $A/B$, one domain occupies one
polar region and one scar appears near the opposite pole [Fig.~\ref{Fig7}(b)].
The remaining nine scars form three latitude bands: one around the equator and
two approximately symmetric bands on either side of it. As $\rho^*$ increases,
the system reaches the most extended coexistence state, ``8 scars + 2 domains''.
This pattern matches the prolate geometry particularly well: two Sq-rich rectangular domains
sit at the two poles, while the eight scars are arranged into two nearly
symmetric off-equatorial bands [Fig.~\ref{Fig7}(c)]. Its large area in the phase
diagram indicates that this two-pole domain arrangement provides an especially
efficient compromise between curvature localization and finite-size packing.

The intermediate ``6 scars + 3 domains'' state occupies the smallest region of
the phase diagram. It is found mainly under oblate conditions, where three
domains are distributed near the equatorial region and the six scars remain more
evenly dispersed outside the domains [Fig.~\ref{Fig7}(d)]. This configuration is
less compatible with the dominant ellipsoidal symmetry and therefore appears
only in a narrow parameter window. By contrast, the ``4 scars + 4 domains''
state again occupies a substantial region, mainly at higher $\rho^*$ on prolate
ellipsoids. In this case, two domains are placed near each polar region, while
the four scars form a spiral-like arrangement around the equatorial zone
[Fig.~\ref{Fig7}(e)]. Finally, the ``2 scars + 5 domains'' state is favored on
oblate ellipsoids. The five domains form an approximately pentagonal
arrangement around the equatorial belt, highlighted by the dashed guide in
Fig.~\ref{Fig7}(f), while the two residual scars are located in opposite
hemispheres. These representative structures show that the coexistence states
are not arbitrary mixtures of scars and domains; rather, each pattern reflects a
specific compromise among defect charge, curvature localization, and packing
constraints.

\begin{figure}[htbp]
    \centering
    \includegraphics[width=0.8\columnwidth]{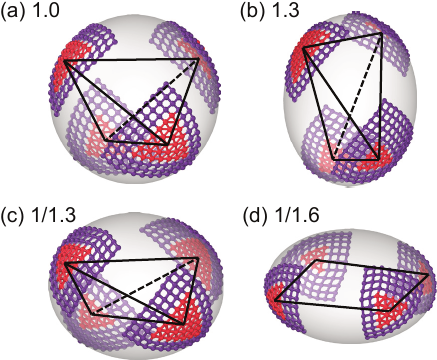}
    \caption{Representative counter-domain configurations corresponding to state (v) in Fig.~\ref{Fig1} at $\rho^*=2.30$ at several aspect ratios $A/B$. (a) Tetrahedral arrangement on the sphere. (b) Elongated arrangement on a
prolate ellipsoid. (c) and (d) Progressively compressed arrangements
on oblate ellipsoids, culminating in localization on the equatorial
belt. Aspect ratio is indicated in each panel.}
    \label{Fig8}
\end{figure}

\textit{Counter domains.}---We use ``counter domain'' for a composite motif in which the local lattice symmetry switches from the surrounding background to that of a host domain and then switches back within the host; the complete motif carries total winding number $w=+1/2$~\cite{xie2025}. Four counter domains form an approximately tetrahedral arrangement on the sphere [Fig.~\ref{Fig8}(a)].

\begin{figure}[htbp]
    \centering
\includegraphics[width=0.6\columnwidth]{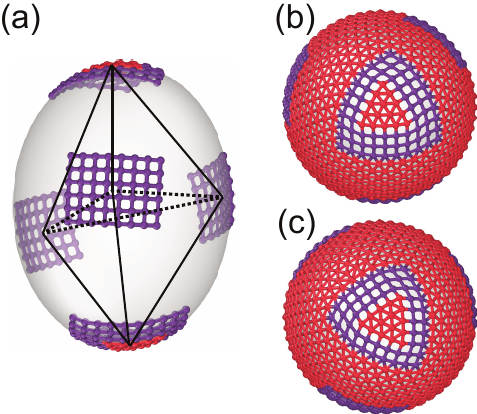}
\caption{Representative coexistence configurations of Sq-rich rectangular domains and counter
domains on a prolate ellipsoid with $A/B = 1.3$ and $\rho^*=2.29$ corresponding to state (iv) in Fig.~\ref{Fig1}. Curvature
inhomogeneity separates the two motifs: counter domains occupy
higher-curvature regions, whereas Sq-rich rectangular domains remain at
lower-curvature latitudes.}
    \label{Fig9}
\end{figure}

Under ellipsoidal confinement, this tetrahedral organization is deformed by the anisotropic curvature field. In prolate ellipsoids [Fig.~\ref{Fig8}(b)], the counter-domain pattern is elongated along the major axis, giving rise to a stretched tetrahedral configuration. In oblate ellipsoids [Fig.~\ref{Fig8}(c)], the same structure is compressed along the symmetry axis and expands preferentially within the equatorial plane. At sufficiently large eccentricity, this confinement becomes strong enough to drive all counter domains onto the equatorial belt, as shown in Fig.~\ref{Fig8}(d). This behavior reflects the coupling between defect charge and curvature: charged vertices are attracted toward curvature extrema, whereas topological conservation and defect-defect repulsion drive a collective reorganization of the entire structure.

\textit{Coexistence of Sq-rich rectangular domains and counter domains.}---Curvature inhomogeneity also permits the two domain types to coexist [Fig.~\ref{Fig9}]. Counter domains preferentially occupy the higher-curvature regions, whereas Sq-rich rectangular domains remain at lower-curvature latitudes. Although a similar spatial arrangement is geometrically possible on a uniformly curved sphere, it is not energetically favored. On the prolate ellipsoid, the curvature gradient stabilizes this spatial separation, allowing the coexistence pattern to emerge as the ground-state structure.

\begin{figure}[htbp]
    \centering
    \includegraphics[width=1\columnwidth]{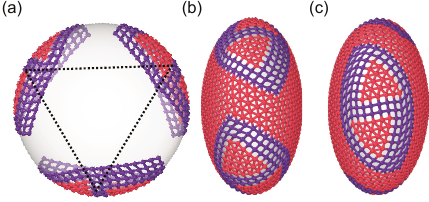}
    \caption{Biangular domains on an oblate ellipsoid corresponding to state (vii) in Fig.~\ref{Fig1} at $A/B=1/1.8$ and $\rho^*=2.35$.
(a) Polar view and (b) and (c) side views of three biangular domains
localized along the high-curvature equatorial belt. The lower-curvature
polar regions remain predominantly Hex ordered.}
    \label{Fig10}
\end{figure}

\textit{Biangular domains.}---A biangular domain extends this nested construction by combining a counter domain with a counter--counter domain, corresponding to a further reversal of local lattice symmetry, giving net winding number $w=+2/3$~\cite{xie2025}. These higher-order structures are also not lowest-energy states on the sphere but appear among the lowest-energy configurations found on strongly oblate ellipsoids. Three biangular domains occupy the high-curvature equatorial belt, while the lower-curvature poles remain largely Hex ordered (Fig.~\ref{Fig10}).

\subsection{Defect patterns in a Sq-dominant background}

\begin{figure}[htbp]
    \centering
    \includegraphics[width=1\columnwidth]{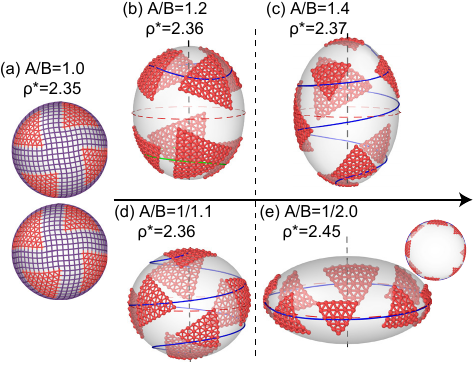}
    \caption{Hex-rich triangular domains in the Sq-dominant regime corresponding to state (ix) in Fig.~\ref{Fig1}. The numerical labels in the panels denote the aspect ratio $A/B$.
(a) Cubic arrangement on the sphere at $A/B=1.0$ and $\rho^*=2.35$. (b) Two four-domain polar rings
at $A/B=1.2$ and $\rho^*=2.36$. (c) Pole-to-pole helical arrangement at $A/B=1.4$ and $\rho^*=2.37$.
(d) Distorted helical arrangement at $A/B=1/1.1$ and $\rho^*=2.36$.
(e) Equatorial ring at $A/B=1/2.0$ and $\rho^*=2.45$.}
    \label{Fig11}
\end{figure}

\begin{figure}[htbp]
    \centering
    \includegraphics[width=0.7\columnwidth]{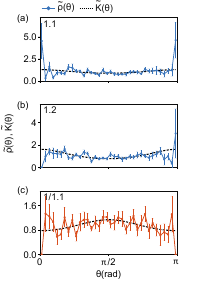}
    \caption{\label{Fig12}
Latitude-resolved comparison of the normalized signed topological-charge density
$\widetilde{\rho}(\theta)$ and normalized Gaussian curvature
$\widetilde{K}(\theta)$ at $\rho^*=2.35$.
Panels (a) and (b) show prolate ellipsoids with $A/B=1.1$ and $1.2$,
respectively, whereas (c) shows an oblate ellipsoid with $A/B=1/1.1$.}
\end{figure}

\begin{figure}[htbp]
    \centering
    \includegraphics[width=1\linewidth]{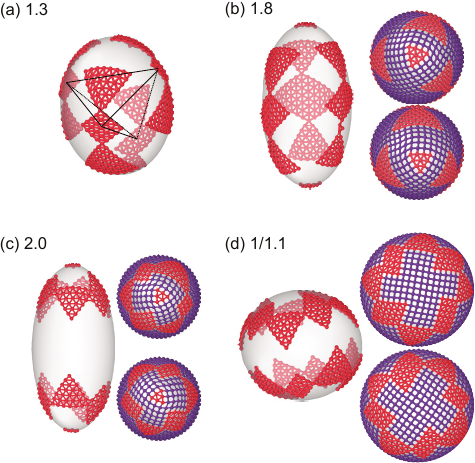}
    \caption{Representative snapshots of bridged states corresponding to state (x) in Fig.~\ref{Fig1}. (a) Four paired-domain complexes at $A/B=1.3$ with $\rho^*=2.40$. (b) Two polar Hex-rich triangular domains and three equatorial complexes at $A/B=1.8$ and $\rho^*=2.41$. (c) Two rotated polar bridging rings at $A/B=2.0$ and $\rho^*=2.43$. (d) Two rotated latitudinal rings at $A/B=1/1.1$ and $\rho^*=2.40$.}
    \label{Fig13}
\end{figure}

\begin{figure}[htbp]
    \centering
    \includegraphics[width=1\columnwidth]{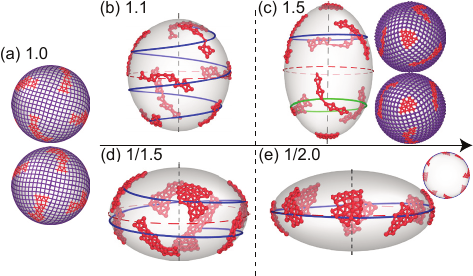}
    \caption{Representative linear scar and open scar configurations corresponding to state (xi) in Fig.~\ref{Fig1}
at $\rho^*=2.70$. (a) Approximately cubic arrangement of open scars on the sphere.
(b) Pole-to-pole spiral distribution at $A/B=1.1$.
(c) Polar open scars surrounded by two rings of linear scars at
$A/B=1.5$. (d) Compressed spiral arrangement at $A/B=1/1.5$.
(e) Equatorial scar ring at $A/B=1/2.0$.}
    \label{Fig14}
\end{figure}

\begin{figure}[htbp]
    \centering

\includegraphics[width=1\columnwidth]{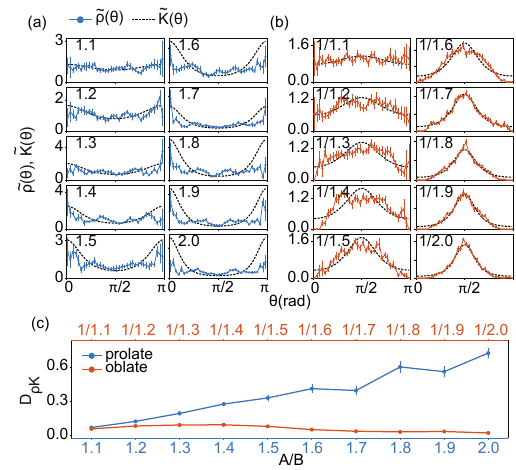}\caption{\label{Fig15}Quantitative defect-curvature comparison for linear and open scars in the Sq-dominant regime at $\rho^*=2.70$. 
(a),(b) Normalized signed topological-charge density
$\widetilde{\rho}(\theta)$ (symbols) and normalized Gaussian curvature
$\widetilde{K}(\theta)$ (dashed curves) for prolate ellipsoid and oblate ellipsoid, respectively.
Symbols and error bars show the mean and standard error over the ten lowest-energy configurations selected from the 20 independent annealing trajectories at each state point.
(c) Defect-curvature deviation index $D_{\rho K}$ versus $A/B$. The mismatch grows strongly in the prolate series but remains small in the oblate series.}
\end{figure}

\textit{Hex-rich triangular domains.}---The Sq-dominant regime contains eight Hex-rich triangular domains, each carrying a net charge of $+1/4$. On the sphere, their centers occupy the vertices of an inscribed cube [Fig.~\ref{Fig11}(a)]~\cite{liu2025transitional}.

Weakly prolate ellipsoids split the domains into two four-domain rings near the poles [Fig.~\ref{Fig11}(b)]. With increasing $A/B$, these rings merge into a pole-to-pole helical arrangement [Fig.~\ref{Fig11}(c)]. Weakly oblate ellipsoids likewise exhibit a distorted helical distribution [Fig.~\ref{Fig11}(d)]. Under stronger flattening, all eight triangular domains migrate toward the equator and form a single ring [Fig.~\ref{Fig11}(e)]. The top-view inset reveals that these equatorial domains assemble into a compass-rose-like pattern.

Figure~\ref{Fig12} provides a quantitative test of curvature selection for Hex-rich triangular domains in the Sq-dominant regime at $\rho^*=2.35$. Only the oblate $A/B=1/1.1$ and prolate $A/B=1.1$ and $1.2$ conditions are included because clean configurations of Hex-rich triangular domains occur over a relatively narrow part of the state diagram. The normalized signed topological-charge density is enhanced toward the equator on the oblate ellipsoid and toward the poles on the prolate ellipsoids, consistent with the locations of the corresponding curvature maxima. These signed topological-charge profiles are nevertheless broader and more strongly fluctuating than the curvature profiles because the charge is carried by a finite number of domain-associated point defects. We therefore use these profiles as a consistency check rather than inferring a systematic deformation trend from the limited data for Hex-rich triangular domains.

\textit{Bridged states.}---A bridged state consists of neighboring Hex-rich triangular domains connected by a parallelogram-shaped Hex bridge. The bridge shares vertices with both domains, changes their relative orientation, and relocates the $+1/12$ point defects from adjacent domain corners to the exposed bridge vertices~\cite{liu2025transitional}. The increased separation between these defects reduces their elastic interaction energy. In this limited sense, the bridged state resembles the VCD state, since both use shared vertices to rearrange point defects and lower the elastic energy, although their local topologies remain distinct. In a VCD contact, two $+1/12$ domain-corner charges are converted into locally neutral vertices with $w=0$, with compensating positive charge appearing elsewhere on the surface. In a bridged motif, by contrast, the $+1/12$ defects are not locally neutralized but are relocated from adjacent triangular-domain corners to the exposed bridge vertices. The fully connected spherical network is not retained on an ellipsoid; only partially bridged states occur within a narrow region of the state diagram.

At $A/B\lesssim1.4$, four paired-domain complexes form an approximately tetrahedral arrangement [Fig.~\ref{Fig13}(a)]. For $1.5\lesssim A/B\lesssim1.8$, two triangular domains occupy the poles and three paired complexes are separated azimuthally by about $2\pi/3$ [Fig.~\ref{Fig13}(b)]. At $A/B>1.9$, each pole is occupied by a triangular domain and surrounded by a ring of three domains and three bridges [Fig.~\ref{Fig13}(c)]. Oblate confinement instead produces two latitudinal bridging rings rotated by approximately $\pi/4$ [Fig.~\ref{Fig13}(d)].

\textit{Linear and open scars.}---At large Sq fraction, a linear scar consists of a sequence of positive and negative winding-number defects with a net charge of $+1/4$~\cite{liu2025transitional}. Despite the shared term, this Sq-background scar is topologically distinct from the five--seven-coordination scar in a Hex-dominant background. As the Hex fraction increases, the chain opens into a Hex-rich corridor bounded by a neutral interface. The morphology changes, but the net charge remains $+1/4$; eight scars therefore supply the required total charge of $+2$ [Fig.~\ref{Fig14}(a)].

On weakly prolate ellipsoids, the scars trace a pole-to-pole path [Fig.~\ref{Fig14}(b)]. Stronger prolateness separates the polar open scars and organizes the remaining defects into two hemispherical rings [Fig.~\ref{Fig14}(c)]. Moderate oblateness retains a distorted pole-to-pole path [Fig.~\ref{Fig14}(d)], whereas strong flattening produces an equatorial ring [Fig.~\ref{Fig14}(e)].

Figure~\ref{Fig15} compares the same normalized defect-charge and curvature
profiles, $\widetilde{\rho}(\theta)$ and $\widetilde{K}(\theta)$, for
linear and open scars in the Sq-dominant background.
Along the prolate series, the curvature maxima progressively contract into narrow polar caps, whereas the defect charge associated with linear and open scars remains distributed over broader latitude ranges. The resulting defect-curvature mismatch therefore increases overall with prolateness. Along the oblate series, both the curvature and the signed topological-charge density remain concentrated around the extended equatorial belt, and the deviation index remains small. The prolate-oblate contrast is thus common to both lattice backgrounds: compact polar curvature maxima promote defect spreading, whereas an extended equatorial maximum allows defects to separate while remaining curvature-localized.

\section{\label{sec:level1-3}Conclusion}

We used simulated annealing with Langevin dynamics to examine how ellipsoidal curvature reorganizes competing Hex and Sq lattices. Across the lowest-energy configurations found for $N=1000$, topology fixed the net defect charge, density selected the predominant lattice background and available motifs, and curvature anisotropy was associated with changes in their positions and connectivity.

In both lattice backgrounds, scar distribution approximately followed the Gaussian-curvature profile under weak deformation. With increasing prolateness, Gaussian curvature became confined to narrow polar caps while scars extended toward lower-curvature latitudes, consistent with competition among curvature localization, finite motif size, and elastic repulsion. Under oblateness, the broad high-curvature equatorial belt allowed scars to remain near the curvature maximum while separating azimuthally.

The prolate--oblate asymmetry therefore arises from the spatial extent of the high-curvature region. Compact polar caps on prolate ellipsoids force finite defect motifs to spread toward lower-curvature latitudes or change their connectivity, whereas the extended equatorial belt on oblate ellipsoids permits azimuthal separation without leaving the high-curvature region.

Domain contact provided a second mode of reorganization. In Hex-dominant configurations, VCD formation locally neutralized contacted $+1/12$ corner charges and was accompanied by compensating positive defects away from the poles. In Sq-dominant configurations, bridges relocated $+1/12$ point defects to the exterior vertices of composite motifs. 

In previous studies, point defects and scars in single-phase lattices accommodated curvature modulation in a straightforward manner~\cite{Irvine2010,LopezJimenez2016,Burke2015}. In the coexisting Hex-Sq system, by contrast, defects also take the form of finite domains with intrinsic shape preferences and characteristic topological-charge patterns. Their interplay with nonuniform curvature produces a much richer repertoire of configurations. The principal distinction from spherical confinement is therefore the emergence of a connectivity response in addition to defect relocation. Nonuniform curvature provides a geometric lever to control both the spatial distribution of topological charge and the connectivity of finite defect motifs in competing lattices.

Because the spreading-versus-separation mechanism originates from curvature localization, finite motif size, and elastic defect interactions, we expect its qualitative form to persist for other soft interactions that support competing Hex and Sq order. The precise density windows, transition aspect ratios, and stability of individual motifs are nevertheless potential-dependent.

These predictions could be tested using soft colloidal or microgel particles confined to shape-controlled curved interfaces. Varying the aspect ratio at approximately fixed surface area and particle number should reveal three observable signatures: defect spreading away from compact polar caps on prolate surfaces, the onset of domain contact or bridging, and azimuthally separated defect rings within the extended equatorial high-curvature belt of oblate surfaces.

\section{Acknowledgement.} This work was financially supported by the National Natural Science Foundation of China (12275137) and Natural Sciences and Engineering Council of Canada. We thank the Digital Research Alliance of Canada for providing computational
resources. We thank Zheng Wang for technical help.

\section{Author Contributions.} Y. L. designed the project. W. L. performed the MD
simulations. B. L., J. Z. Y. C., and Y. L. supervised the project. All authors are involved in writing and revising the manuscript.

\section{Competing interests.} The authors declare no competing interests.

\section{Data Availability}
The data that support the findings of this study are available from the corresponding author upon reasonable request.
\clearpage
\onecolumngrid
\appendix

\section{Local structures and charge assignments}
\label{app:defect-atlas}

To make the structural distinctions in the state diagram directly
accessible, Fig.~\ref{FigA1} provides a uniform local-view atlas of all
states identified in Fig.~\ref{Fig1}.

\begingroup

\begin{figure*}[!ht]
    \centering
    \includegraphics[
        width=\textwidth,
        height=0.62\textheight,
        keepaspectratio
    ]{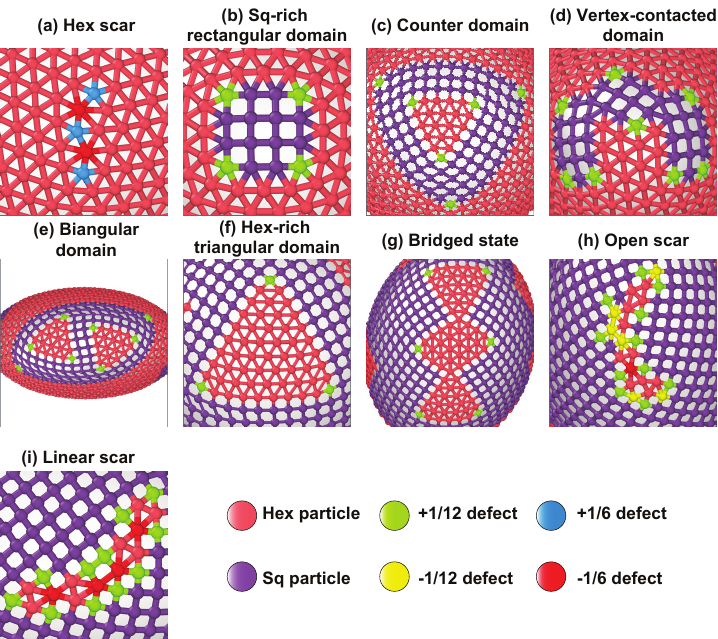}
    \caption{
    Local atlas of the defect states identified in Fig.~\ref{Fig1}.
    Each panel presents an enlarged local view of the bond network,
    highlighting the characteristic local structure, domain contacts,
    and charge arrangement of the corresponding state.
    Particle colors encode both the local lattice environment and the
    local topological charge (winding number) $w$.
    Neutral particles ($w=0$) in locally Hex- and Sq-ordered environments
    are shown in pink and purple, respectively.
    Green, yellow, blue, and red particles carry topological charges
    $w=+1/12$, $-1/12$, $+1/6$, and $-1/6$, respectively.
    }
    \label{FigA1}
\end{figure*}

\endgroup

\clearpage
\twocolumngrid

\end{document}